\documentclass[%
 preprint,
superscriptaddress,
 amsmath,amssymb,
 aps,
 prd
]{revtex4-2}

\usepackage{graphicx}
\usepackage{dcolumn}
\usepackage{bm}
\usepackage{soul}
\usepackage[final]{changes}
\usepackage[mathlines]{lineno}
\switchlinenumbers

\begin{document}
\preprint{}
\title{Search for \added{relativistic }fractionally charged particles in space}


\author{F.~Alemanno}
\affiliation{Gran Sasso Science Institute (GSSI), Via Iacobucci 2, I-67100 L'Aquila, Italy}
\affiliation{Istituto Nazionale di Fisica Nucleare (INFN) -Laboratori Nazionali del Gran Sasso, I-67100 Assergi, L'Aquila, Italy}

\author{C.~Altomare}
\affiliation{Istituto Nazionale di Fisica Nucleare, Sezione di Bari, I-70126, Bari, Italy}

\author{Q.~An}
\affiliation{State Key Laboratory of Particle Detection and Electronics, University of Science and Technology of China, Hefei 230026, China}
\affiliation{Department of Modern Physics, University of Science and Technology of China, Hefei 230026, China}

\author{P.~Azzarello}
\affiliation{Department of Nuclear and Particle Physics, University of Geneva, CH-1211, Switzerland}

\author{F.~C.~T.~Barbato}
\affiliation{Gran Sasso Science Institute (GSSI), Via Iacobucci 2, I-67100 L'Aquila, Italy}
\affiliation{Istituto Nazionale di Fisica Nucleare (INFN) -Laboratori Nazionali del Gran Sasso, I-67100 Assergi, L'Aquila, Italy}

\author{P.~Bernardini}
\affiliation{Dipartimento di Matematica e Fisica E. De Giorgi, Universit\`a del Salento, I-73100, Lecce, Italy}
\affiliation{Istituto Nazionale di Fisica Nucleare (INFN) - Sezione di Lecce, I-73100, Lecce, Italy}

\author{X.~J.~Bi}
\affiliation{University of Chinese Academy of Sciences, Beijing 100049, China}
\affiliation{Particle Astrophysics Division, Institute of High Energy Physics, Chinese Academy of Sciences, Beijing 100049, China}

\author{M.~S.~Cai}
\affiliation{Key Laboratory of Dark Matter and Space Astronomy, Purple Mountain Observatory, Chinese Academy of Sciences, Nanjing 210023, China}
\affiliation{School of Astronomy and Space Science, University of Science and Technology of China, Hefei 230026, China}

\author{E.~Casilli}
\affiliation{Dipartimento di Matematica e Fisica E. De Giorgi, Universit\`a del Salento, I-73100, Lecce, Italy}
\affiliation{Istituto Nazionale di Fisica Nucleare (INFN) - Sezione di Lecce, I-73100, Lecce, Italy}

\author{E.~Catanzani}
\affiliation{Istituto Nazionale di Fisica Nucleare (INFN) - Sezione di Perugia, I-06123 Perugia, Italy}

% ...

\collaboration{DAMPE Collaboration}\email{Electronic address: dampe@pmo.ac.cn}



\date{\today}

\begin{abstract}
More than a century after the performance of the oil drop experiment, the \added{possible} existence of fractionally charged particles (FCPs) still remains \replaced{unsettled.}{a mystery in modern physics.} The search for FCPs is crucial for \replaced{some}{the} extension\added{s} of the Standard Model in particle physics. Most of the previously conducted \replaced{searches for}{studies on} FCPs in cosmic rays were based on \deleted{observational}experiments underground or at high altitudes. 
However, \replaced{there  have been few searches for FCPs in cosmic rays carried out \replaced{in}{on} orbit} {there is rarely experiment to study FCP from cosmic rays on orbit} \replaced{other than}{except the} \deleted{experiments}\replaced{AMS-01 flown by}{of carrying on} a space shuttle \replaced{and BESS by \added{a }balloon at the top of}{and flying on a balloon near} the atmosphere. In this study, we conduct \added{an} FCP search in space based on on-orbit data obtained using the DArk Matter Particle Explorer (DAMPE) satellite over a period of five years. Unlike underground experiments, which require \added{an} FCP energy of the \replaced{order}{magnitude} of hundreds of GeV, our FCP search \replaced{starts at}{is initiated from} only a few GeV. An upper limit of $6.2\times 10^{-10}~~\mathrm{cm^{-2}sr^{-1} s^{-1}}$ is \replaced{obtained}{considered} for the flux. 
Our results demonstrate that \replaced{DAMPE}{the proposed method} exhibits higher sensitivity than experiments of similar types by three orders of magnitude\replaced{ that}{, which also} more stringently restricts the conditions for the existence of FCP in primary cosmic rays.
\newpage

\end{abstract}

\maketitle

\section{\label{sec:level1}Introduction}

The oil drop experiment \deleted{was }originally performed by Robert A. Millikan in 1909 \cite{millikan1909new} \replaced{provided}{. It achieved} the first direct measurement of the electric charge ($e$) of an electron. Since then, \replaced{particles detected}{all detected particles} have been observed to \replaced{carry}{have} charges that are \added{integer }multiples of $e$. It was believed that $e$ is the smallest charge in nature\deleted{It was not} until Gell-Mann and Zweig proposed the quark model in 1964 ~\cite{gell1964schematic, zweig1964_3} \replaced{that states}{that people realized that} quarks, as elementary particles, carry fractional charge values ($\frac{1}{3}e$ or  $\frac{2}{3}e$). \replaced{Thus, until theories of quantum chromodynamics (QCD) indicated that free quarks do not exist, research interest in fractionally charged particles (FCPs) was driven by the search  for free quarks.}{Once a while, research interest in fractionally charged particles (FCPs) was driven by the search  for free quarks. However, quantum chromodynamics (QCD) indicates that free quarks do not exist.}
\replaced{However, benefitting}{Benefit} from the development of \added{new} theories, FCPs are allowed in some extensions of\added{ the} Standard Model (SM). For example, \deleted{the }extensions \replaced{of}{to} \added{the} SM gauge group SU(5) predict \replaced{a color}{the colour} singlet particle \added{with a charge} of $\frac{1}{3}e$~\cite{1982Fractionally, S1983Fractional}. In some larger groups, the existence of FCPs arise\added{s} from a natural derivation~\cite{1984SU, 1983Fractionally, 1983Fractional}.
 Since then, \deleted{research in }this \replaced{field of study}{field} \replaced{has been}{was} oriented towards the search for any new \replaced{fractionally charged particle}{particle with fractional charge}.

FCPs are assumed to be a type of heavy lepton~\cite{langacker2011requiem}\replaced{, that have a }{. They are expected to exhibit high }penetrating ability and \replaced{are}{be} free from high energy cascade effects\replaced{ other than}{, except} ionization and weak interaction\added{s}. 
\replaced{Over the last few decades, several experimental studies have been conducted to find counterexamples to the QCD assertion of the non-existence of free quarks and to directly search for new particles with fractional charge.}{Several experimental studies have been conducted over the last few decades both to find counterexamples to the QCD assertion of the non-existence of free quarks and simply search for new particles with fractional charge.} There are three primary methods to detect FCPs  ~\cite{perl2004brief}. 
\added{First, there is the}\deleted{The} modern Millikan oil drop technique\replaced{ that utilizes}{which is utilized in} an oscillating electric field  ~\cite{smith1989searches, halyo2000search} to search for \replaced{contained FCPs}{FCPs contained} within the bulk matter \replaced{on the earth}{of earth materials}. \added{Second, }high-energy accelerators are used to search for FCPs \replaced{in the process of particles production }{by simulating the big bang }to test superstring models or explore physics theories beyond the standard model ~\cite{lyons1987search, cecchini1993fragmentation, jones1977review, chatrchyan2013search}. 
\replaced{Last}{Finally}, FCPs are also searched for in cosmic rays, and such a method can be subclassified into the following three types.  1) \replaced{Ground-based}{Plateau and sea level} experiments ~\cite{mccusker1969evidence, hsiao1982search} are \replaced{detecting}{conducted to detect} cores of extensive air showers. \added{In 1969, }McCusker and Cairns \replaced{claimed}{declared} the discovery of a free $\frac{2}{3}e$ particle in a cloud chamber image  ~\cite{mccusker1969evidence} of an extensive air shower\deleted{ using the aforementioned method}, however, \replaced{no replication was ever achieved.}{no provable evidence was supplied.} 2) Underground experiments ~\cite{mori1991search, aglietta1994search, ambrosio2000search, ambrosio122004final, agnese2015first, alvis2018first} \deleted{are conducted  to }\replaced{evade}{exclude} background noise from extensive air showers and attempt to observe FCPs that pass through \replaced{the}{mountainous and flat} overburden\deleted{s}. Such FCPs would have to \replaced{start out with}{possess} \added{an} energy larger than hundreds of GeV to penetrate \deleted{terrestrial }rocks before entering the underground laboratory. 
\replaced{With a}{Concerning the} large acceptance and long exposure time, the underground experiment MACRO \replaced{obtained}{released} a flux upper limit of $6.1\times 10^{-16}~~\mathrm{cm^{-2}sr^{-1} s^{-1}}$ at\added{ the} 90\% confidence level (C. L.) ~\cite{ambrosio122004final} for particles with charges from $\frac{1}{4}e$ to  $\frac{2}{3}e$. 3) \replaced{Searches for FCPs in cosmic rays are}{The search for FCPs is} also conducted in \deleted{cosmic rays in upper }space, notably, on \added{the} space shuttle (AMS-01 ~\cite{sbarra2003search}) and balloon (BESS ~\cite{fuke2008search}). Compared to underground experiments, particles with significantly lower energy in the order of a few GeV\deleted{,} \replaced{are able to}{can} be observed \replaced{in space experiments}{using experiments in space}\replaced{ where}{. From the two aforementioned experiments,} the stricter flux upper limit of  $3.0\times 10^{-7}~~\mathrm{cm^{-2}sr^{-1} s^{-1}}$   for FCPs was obtained from AMS-01. However, both \added{the }BESS and AMS-01 experiments were short-lived, and there has been no long-term continuous search for FCPs based on on-orbit experimentation.

\section{\label{sec:level1}DAMPE mission}

\begin{figure}[!htb]
	\includegraphics[width=\linewidth]{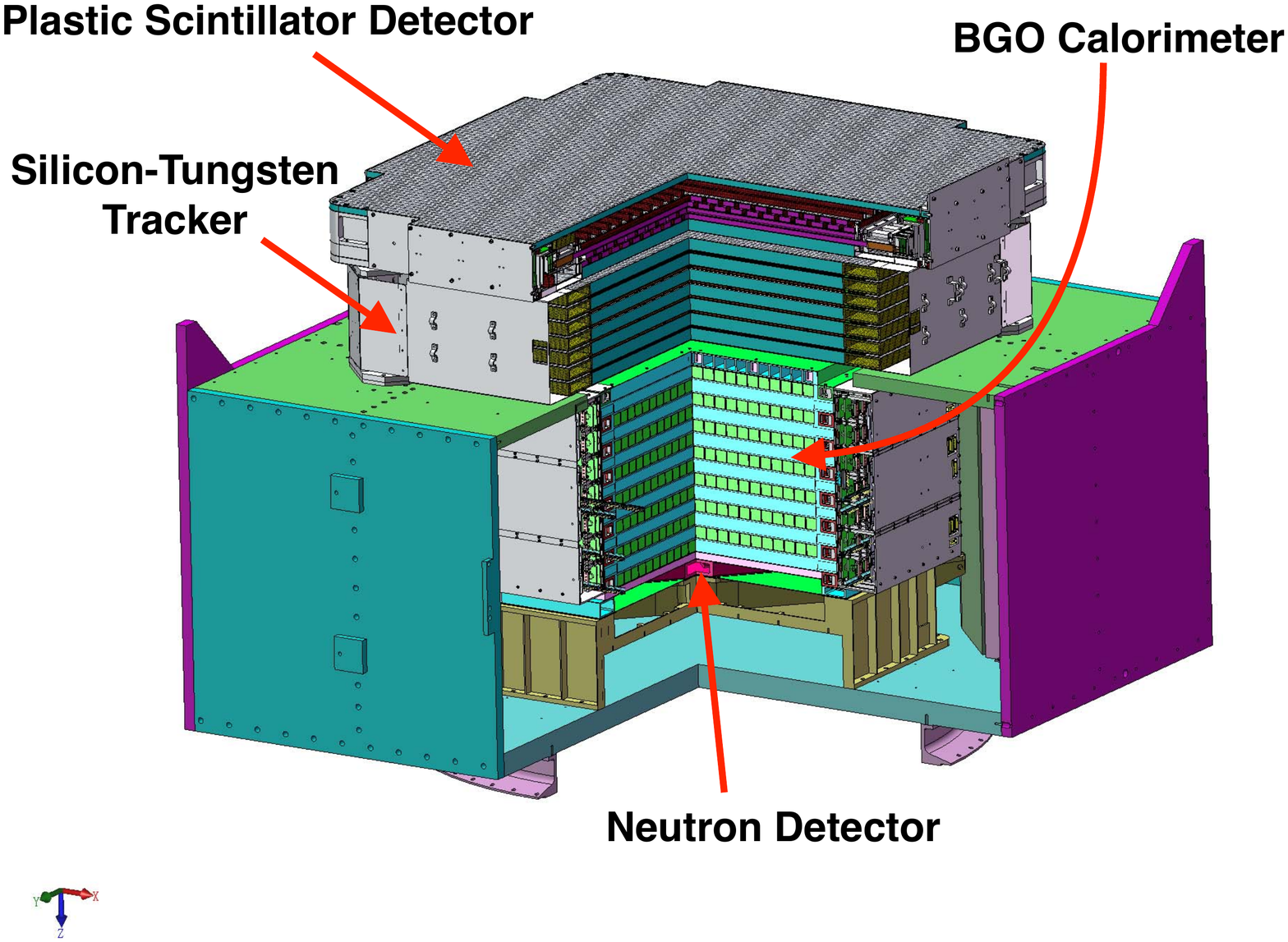}
	\caption{\label{fig4} \added{The structure of the DAMPE detector.}}
\end{figure}

\added{The DArk Matter Particle Explorer (DAMPE ~\cite{chang2017dark, ambrosi2019orbit}, also known as "Wukong" in Chinese) is an on-orbit calorimetric-type, satellite-borne detector that can be used to search for FCPs in primary cosmic rays in space.}
\deleted{In this context, the DArk Matter Particle Explorer (DAMPE, also known as "Wukong" in China) is a suitable on-orbit apparatus that may be used to search for FCPs in primary cosmic rays in space. DAMPE is a calorimetric-type, satellite-borne detector.} 
\added{One of the scientific objectives of DAMPE is to search for dark matter in an indirect approach that involves examining high-energy cosmic rays in space.}
\deleted{One of the scientific objectives of DAMPE is to observe high-energy cosmic rays in space directly to facilitate the indirect search for dark matter.}
\replaced{Launched on 17 December 2015, the DAMPE detector has been in stable operation on the 500 km Sun-synchronous orbit for more than six years,}{The DAMPE detector was launched into a 500 km Sun-synchronous orbit on 17 December 2015. It has been in stable operation on orbit for more than six years,} and \added{has} made important contributions to \added{the }cosmic ray observation of electrons~\cite{ambrosi2017direct}, protons~\cite{an2019measurement}, and helium~\cite{alemanno2021measurement}. 
From top to bottom, DAMPE consists of \added{four sub-detectors as shown in} Fig.~\ref{fig4}. \replaced{A}{a} Plastic Scintillator Detector (PSD) ~\cite{yu2017plastic, ma2019method}\added{, that is made of two layers of 41 plastic scintillation strips.} \replaced{A}{a} Silicon-Tungsten tracKer converter (STK) ~\cite{azzarello2016dampe, tykhonov2018internal},\added{ that is composed of six modules of double-layer silicon-strip detectors, with three layers of tungsten inserted.} \replaced{A}{a} Bismuth Germanium Oxide (BGO) imaging calorimeter ~\cite{zhang2016calibration}\added{ consisting 14 layers of 22 crystal bars, corresponds to $\sim$ 32 radiation lengths and 1.6 nuclear interaction lengths.} \replaced{It also includes}{And} a NeUtron Detector (NUD) ~\cite{huang2020calibration}\added{, that is a collection of four boron-doped plastic scintillation tiles}.
The PSD measures \added{the }charge of \added{the} incident particle and contributes to the anticoincidence measurement of gamma rays. 
The STK reconstructs the trajectory and also measures \replaced{the particle's charge}{their charge}. The BGO calorimeter measures the energy\replaced{,}{ and} provides particle identification, and\deleted{ also} provides the trigger \replaced{for}{of} the DAMPE spectrometer. The NUD provides additional electron/hadron discrimination.
\added{The PSD strips and BGO crystals of DAMPE are arranged in parallel in each layer and orthogonally arranged between adjacent layers.}
The YOZ plane\added{s} and XOZ planes are defined as lateral views perpendicular to the odd and even layers of \added{the }BGO calorimeter (numbered from 1 to 14), respectively.

DAMPE has good charge resolutions of 0.06$e$ and 0.04$e$ for\added{ measuring singly charged particles with} the PSD ~\cite{dong2019charge} and STK ~\cite{vitillo2017measurement}, respectively. Further\added{more}, compared to similar types of space experiments, DAMPE has a relatively large\deleted{r} acceptance and long\deleted{er} exposure duration, which are advantageous in search\added{ing} for FCPs. 
\replaced{Here}{Thus, in this study,} we conduct\deleted{ed} \replaced{a}{the} search for FCPs based on on-orbit data \replaced{collected with}{obtained using} DAMPE over a period of five years.

\section{\label{sec:level1}Data analysis}
\subsection{\label{sec:level2}Target FCPs}
The on-orbit \replaced{data}{sample} corresponding to the latitude region of [-20$^\circ$, +20$^\circ$] is used to search for FCPs\deleted{ because of the high energy threshold geomagnetic cutoff of charged particles. }\added{, where the  strength of geomagnetic field is generally uniform,} \replaced{and the energy cutoff}{which} is usually $\sim$ 10 GeV for singly charged particles. 
\replaced{Given}{Considering} \added{that} the acceleration mechanism of cosmic rays may \added{be }related to \replaced{their}{its} charge~\cite{alemanno2021measurement}, \replaced{FCPs carry proportionately lower energy}{the energy of FCPs is probably not very high. Thus, it is necessary to search for low energy FCPs}. 
\replaced{Combined with}{By} the heavy lepton assumption, the search for FCP\added{s} is constrained to Minimum \replaced{Ionizing}{Ionisation} Particle\added{s} (MIP\added{s}).
\added{While particles above the threshold primarily lie in the  relativistic energy region, charge measurements  using both PSD and STK are in approximate agreement with the Bethe-Bloch equation. Based on the measurements, the}
\deleted{Particles above the threshold primarily lie in the relativistic energy region, where charge measurements using both PSD and STK are in approximate agreement with the Bethe-Bloch principle. The} reconstructed charge value is the “\replaced{k}{K}ey” information for the FCP detection. 
The measurement of energy deposition is expressed \replaced{in units of the energy deposited by a}{as MIPs for} singly charged MIP event\deleted{s}, \replaced{which}{where a MIP event} deposits approximately 23 MeV in one BGO crystal ~\cite{zhang2016calibration}. Compared with \replaced{a}{the} singly charged MIP event, the energy deposition of \added{an} FCP is \replaced{proportional to}{in proportion with} \replaced{the squared value of the particle's charge.}{the squared charge value.} 
The trigger system \replaced{is generated by}{is based on} the BGO calorimeter, whose threshold for a MIP event is calibrated to be approximately $\frac{1}{5}$ MIP\deleted{s} ~\cite{zhang2019design} based on on-orbit data. \replaced{Thus, due to the very low trigger efficiency for FCPs with $\frac{1}{3}e$ ($\frac{1}{9}$ MIP)}{It has low efficiency in the detection $\frac{1}{3}e$ ($\frac{1}{9}$ MIPs) FCPs. As a result}, this study focuses on $\frac{2}{3}e$ \deleted{($\frac{4}{9}$ MIPs)}FCPs. 

\subsection{\label{sec:level2}Background estimation}
Due to the limited charge resolution, high-energy protons/antiprotons, electrons/positrons, and high energy gamma\replaced{~}{-}rays are the primary sources of background noise.
The BGO calorimeter \replaced{is}{has} approximately 32 radiation lengths \added{deep}\replaced{, thus excluding misidentifications caused by electrons/positrons and gamma rays}{. Therefore, all electrons/positrons and gamma-rays produce electromagnetic showers in it and misidentifications arising from it can be ignored}. \replaced{Moreover, the 1.6 nuclear interaction lengths deep such that 80\% of protons/antiprotons develop hadronic showers; therefore, misidentification from the 20\% non-showering, MIP-like high-energy protons/antiprotons is the largest source of background.}{Approximately 80\% of protons/antiprotons exhibit hadronic showers in the BGO calorimeter as the nuclear interaction length is 1.6. Hence the misidentification of high-energy protons/antiprotons that do not exhibit hadron interaction in the BGO calorimeteris the largest source of background radiation.}

\subsection{\label{sec:level2}Monte Carlo simulations}
 Monte Carlo (MC) \replaced{simulations of}{simulated} protons and FCPs with $\frac{2}{3}e$ based on the GEANT4 ~\cite{agostinelli2003geant4} are used to study the background and signals. 
 \replaced{GEANT4 is capable of performing simulations on (virtual ) particles with selected mass, charge, and physical process.}{GEANT4 provides an interface for adding a new (virtual) particle as desired. The charge, mass, and physical processes are all optional for the virtual particle.} \added{Thus, we} \deleted{We }\replaced{insert a virtual}{created a} MIP-like FCP with $\frac{2}{3}e$ within the GEANT4 framework in the DAMPE software. Since the energy deposition of relativistic heavy lepton-like particles with\added{ a} certain charge value is nearly independent of their mass\added{es}, the mass of \added{the} FCP \replaced{i}{wa}s arbitrarily taken to be 1200 MeV. The original index of the energy spectrum \replaced{i}{wa}s taken to be -2.7, \replaced{in agreement with the all-particle cosmic ray spectrum}{in consonance with the nature of cosmic rays}. The processes of multi\added{ple} scattering and ionization \replaced{a}{we}re added. The sample of MC FCP\added{s} \replaced{i}{wa}s used as the signal to be analyzed. Both primary and secondary protons \replaced{a}{we}re taken into account \replaced{in evaluating the}{with respect to} background\deleted{ radiation}.

\subsection{\label{sec:level2}Event selections}

The MIP events are selected during the search for FCPs. The detailed event selection method is described below.

\begin{itemize}
	\item \textit{Trigger efficiency.} DAMPE comprises four trigger patterns ~\cite{zhang2019design}\added{:} Unbiased Trigger (UNBT), MIPs Trigger (MIPT), High Energy Trigger (HET), and Low Energy Trigger (LET).
	\added{The UNBT and MIPT are used in the event selection. The UNBT requires that the energy deposition of each of the first two BGO layers to be larger than $\frac{1}{5}$ MIP. The MIPT has two modes to select events that penetrate the BGO calorimeter from top to bottom. One requires the energy deposition of the layers 1, 11, 13 to be larger than $\frac{1}{5}$ MIP, while the other requires that of layers 2, 12, 14 to be larger than $\frac{1}{5}$ MIP.}
	To reduce the \replaced{amount of data}{stress of} storage, different pre-scale factors are applied independently to the \replaced{UNBT}{UBNT}, MIPT, and LET events \replaced{during}{corresponding to} operation at different latitudes. \replaced{UNBT}{UBNT} has a relatively loose \replaced{restriction}{limit} and is used to estimate the efficiency of the other triggers.\\
	
	\replaced{Based on the on-orbit data, }{The MIP events are selected from on-orbit data, and} the MIPT is observed to be active only \replaced{in}{over} the latitude range, [-20$^{\circ}$, +20$^{\circ}$], where the pre-scale factors for MIPT and UNBT are $\frac{1}{4}$ and $\frac{1}{512}$, respectively. The\deleted{ calculation of the} trigger efficiency of on-orbit data requires \added{the }consideration of the \deleted{following }pre-scale factor\added{s and is given by}:
	\begin{eqnarray}
		\epsilon_{MIPT} = \frac{N_{MIPT}\times 4}{N_{UNBT}\times 512}
		\label{eq2}\replaced{,}{.}
	\end{eqnarray}

	where $N_{MIPT}$ denotes the number of \replaced{prescaled MIPT events}{events with MIPT after applying the $\frac{1}{4}$ factor}, \added{and }$N_{UNBT}$ is the number of \replaced{prescaled UNBT events}{events with UNBT after applying the $\frac{1}{512}$ factor}. The efficiency is observed to be generally steady at 97.3\% as a function of time. No pre-scale factor is applied to the MC sample, and the trigger efficiencies \replaced{for}{of} MC protons and FCPs \added{of $\frac{2}{3}e$} are 99.4\% and 85.5\%\added{,} respectively.

	\item \textit{Track selection.}\added{~}BGO tracks are reconstructed for each event passing through the BGO calorimeter. Ideally, the BGO track is required to be completely contained within the detector while implementing \added{a} geometr\replaced{ic}{y} selection. Moreover, coincidence of the track with the edges of the first layer \added{of} PSD strips or the edges of the last layer \added{of} BGO crystals \replaced{are}{should be} \replaced{excluded}{avoided}. \replaced{Then}{Thus}, a satisfactory track reconstruction using \added{the} STK \replaced{with}{and} the track seed provided by \added{the} BGO is required  ~\cite{tykhonov2017reconstruction}. In order to evaluate the track efficiency, a sample is selected based on the BGO track, and the efficiency is estimated \replaced{from}{by} \deleted{calculating }the \replaced{fraction}{proportion} of the sample selected using the STK track. Thus, the track efficiency is given by
	\begin{eqnarray}
	\epsilon_{track} = \frac{N_{STK\&BGO}}{N_{BGO}}
	\label{eq2}\replaced{,}{.}
    \end{eqnarray}

	where $N_{BGO}$ denotes the number of events selected with the BGO track and $N_{STK\&BGO}$ is the number of further events corresponding to the STK track.\\
	
	\begin{figure*}[!htb]
		\includegraphics[width=0.8\linewidth]{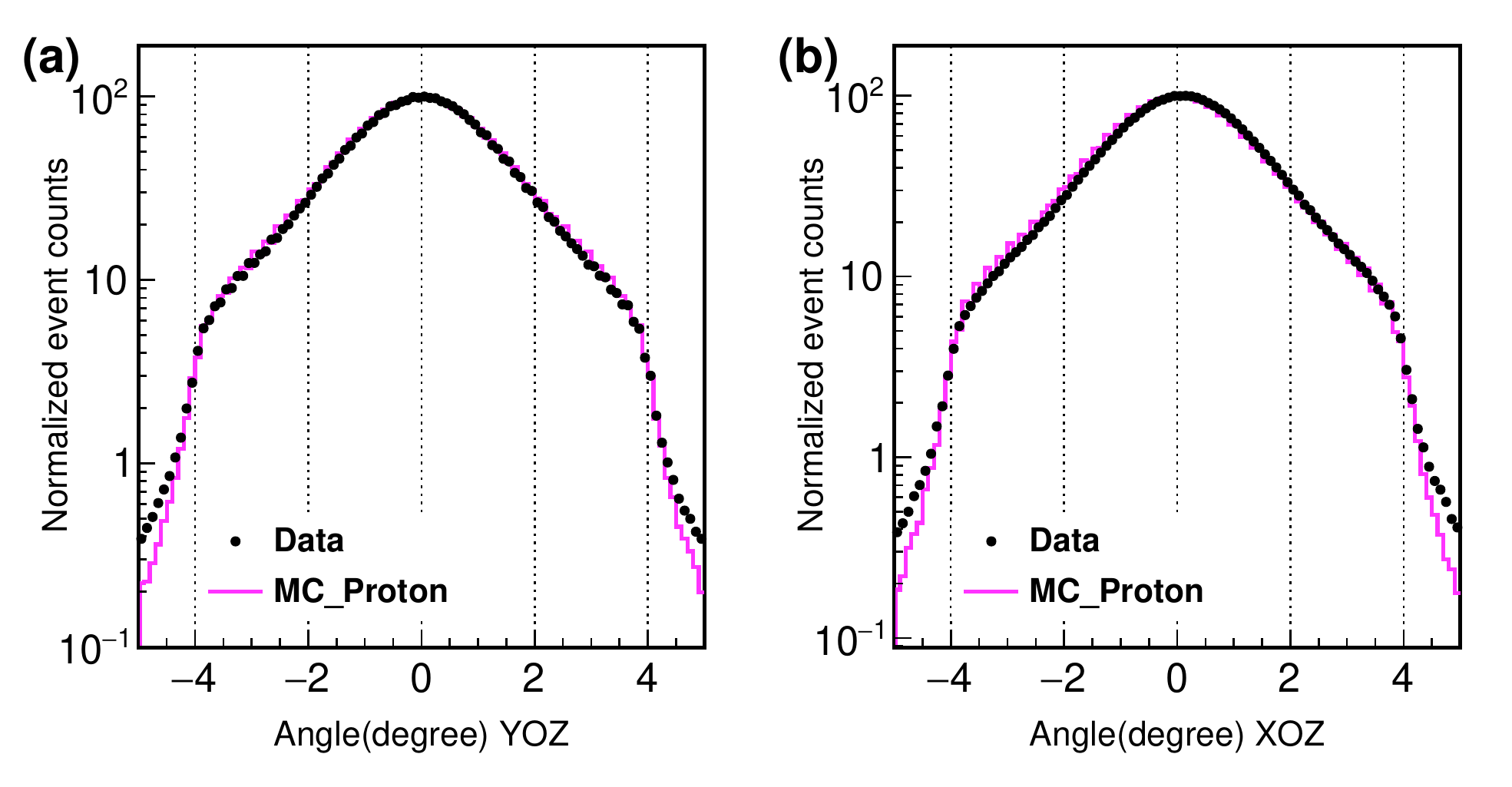}
		\caption{\label{figs1} The angle difference between the BGO track and STK track. The (a) and (b) \added{panels} depict the distributions of angle difference in the YOZ and XOZ planes, respectively. \replaced{The}{It can be seen that the} angle differences obtained from the on-orbit data (black dot\added{s}) and MC proton (pink solid line) agree well within [-4$^{\circ}$, +4$^{\circ}$].}
	\end{figure*}

	Considering the large size ($25~\mathrm{mm}\times 25~\mathrm{mm}\times 600~\mathrm{mm}$) of \replaced{each}{a} BGO crystal, the reconstruction of a BGO track has \added{a} relatively large\deleted{r} uncertainty. The angle difference between the BGO track and STK track is restricted\added{ in order} to \replaced{eliminate}{ignore} scattering events and enhance the reliability of the STK track. 
	\deleted{ The PSD strips and BGO crystals of DAMPE were arranged in parallel in each layer and orthogonally arranged between adjacent layers. 
	The YOZ plane\added{s} and XOZ planes were defined as lateral views perpendicular to the odd and even layers of BGO calorimeter (numbered from 1 to 14), respectively.}
	 As depicted in Fig.~\ref{figs1}, the distributions of \added{the }angle difference between \added{the }on-orbit data and MC protons exhibit good agreement \replaced{in}{over} [-4$^{\circ}$, +4$^{\circ}$] \replaced{in}{on} both the YOZ and XOZ planes. The discrepancy between \added{the MC proton and on-orbit data in}\deleted{ at} the region of larger angle\added{s} \deleted{ difference where the events are removed }can be attributed to \replaced{multiple~}{multi-}scattering. 
	
	\item \textit{MIP requirements.} To purify the MIP events, the respective MIP requirements are  \replaced{applied to}{organized for} \added{the} PSD and \added{the} BGO. Besides the fired PSD strip and BGO crystal, some events may be oblique injections through adjacent cells of strips or crystals, or \replaced{generating}{evolving} knock-on electrons. In such cases, multiple fired cells appear on the PSD and BGO layers. Thus, a relatively loose limit of 2 strips and 2 crystals \added{per layer} is applied to aid the selection of MIP events.
	In \added{the} PSD, at most two fired strips are \replaced{allowed}{required} in each layer. Since the energy fluctuation\added{s} induced by each event passing through a corner of the PSD strip significantly affect\deleted{s} the charge reconstruction, the fired strips \replaced{are}{is} required to be penetrated by the trajectories from the top surface to \added{the} bottom surface. Similarly, in \added{the} BGO calorimeter, at most two fired crystals are \replaced{allowed}{required} in each layer and more than ten layers\added{ of the BGO} are required to have signals. 
	Moreover, at least one of the final two layers\added{ (13 or 14) of the BGO} is required to have a signal to ensure the full penetration \replaced{through}{of an event in} the BGO calorimeter. 
	
	\item \textit{Charge reconstruction.} \replaced{Signals from}{has independent read-outs at} both ends\replaced{ of a fired}{As each} PSD strip  \deleted{, they} should be consistent \replaced{within the}{following} fluorescence attenuation correction. To ensure the reliability of charge reconstruction, the ratio of charge values at the two ends of the PSD strip is constrained. The average value of two PSD layers is taken to be the PSD charge. The STK charge is also taken to be the average of the charge values corresponding to multiple layers after correction ~\cite{vitillo2017measurement}.
    The results of charge reconstruction are depicted in Fig.~\ref{fig1}. MC FCPs and singly charged MIPs are adequately distinguishable in both \added{the} PSD and STK. The charge spectra obtained from the on-orbit data and MC protons display close similarity.

\begin{figure*}[!htb]
	\includegraphics[width=\linewidth]{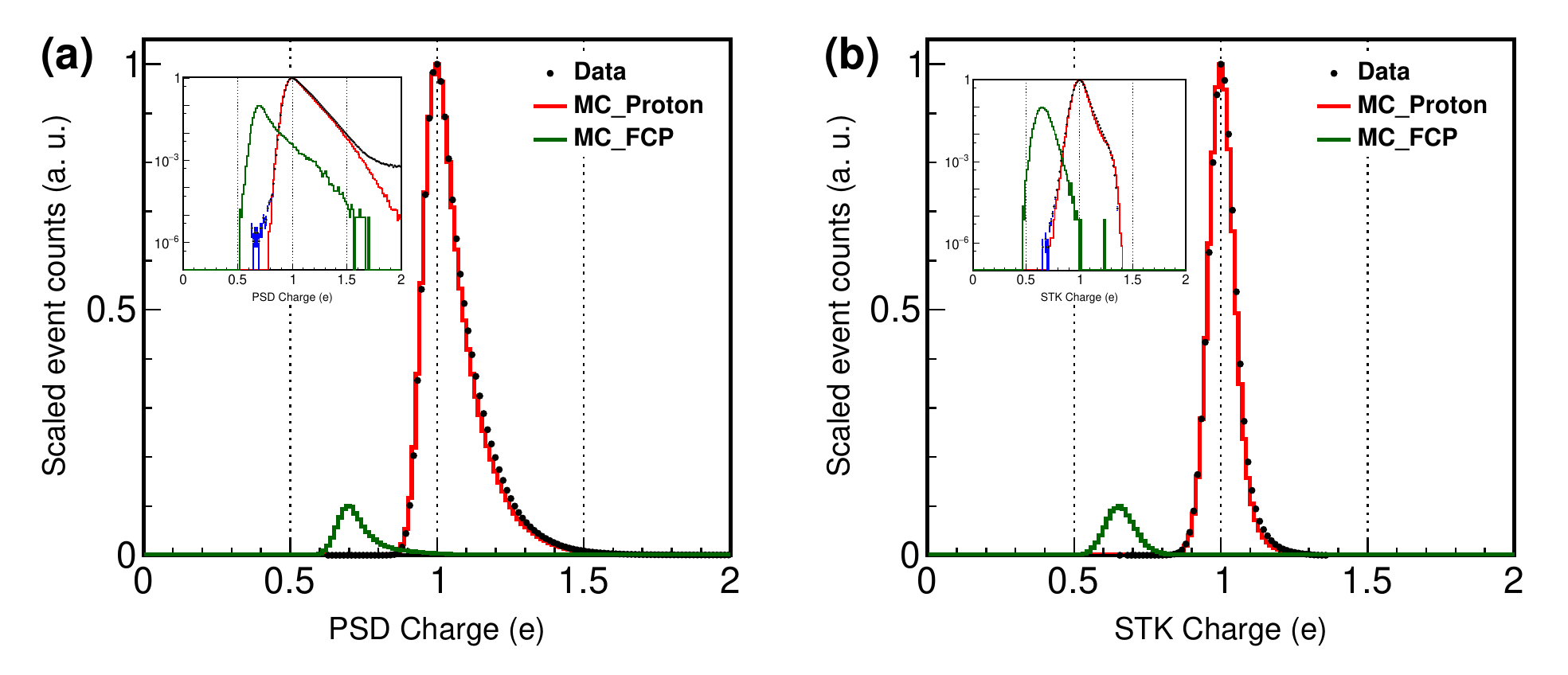}
	\caption{\label{fig1} The distributions of charges measured by \added{the} PSD (a) and STK (b). The spectra obtained from on-orbit data (black dots) and MC protons (red line) are normalized\added{ with ordinate values in arbitrary units} and \added{are }consistent. The FCP (green line) is well distinguished from singly charged particles. The log scale distributions \replaced{for}{of} \added{the} PSD and STK are \replaced{shown in the insets}{posted at the corresponding linear scale figures}.}
\end{figure*}
	
\end{itemize}

\subsection{\label{sec:level2}Definition of \added{the }signal region}
\begin{figure*}[!htb]
	\includegraphics[width=\linewidth]{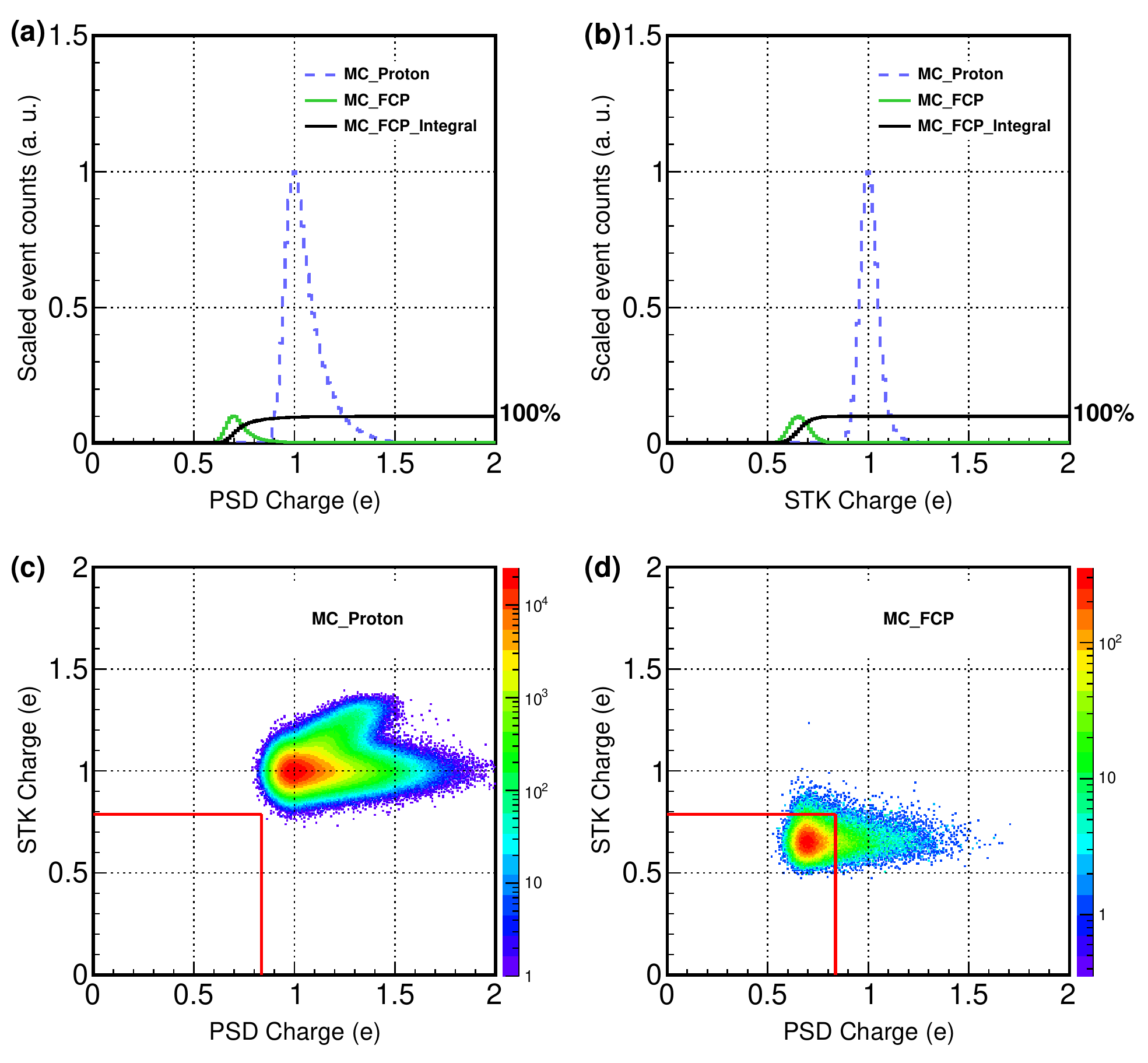}
	\caption{\label{figs2} The charge distributions \replaced{from the}{of} PSD and STK and the definition of the signal region. The charge distributions \replaced{for the}{of} PSD and STK are shown \replaced{in panels}{as} (a) and (b), respectively. \added{The event counts are scaled to arbitrary units. }The solid green lines correspond to MC FCP\added{s} and the dashed \deleted{light} blue lines are for MC proton\added{s}. With the help of the integrals of FCP\deleted{s} charges represented by the solid black lines, the signal region is defined based on the distributions presented in \added{panels} (a) and (b). The red lines in \added{panels} (c) and (d) represent the signal region for FCPs where the charge values for \added{the} PSD and STK are 0.84$e$ and 0.79$e$,  respectively. All background proton MIP events fall outside the region, as depicted in \added{panel} (c). \deleted{The injections from bottom to top are simulated and excluded as well.} Injections from the bottom surface to the top surface of DAMPE are simulated and excluded as well. The combined \replaced{efficiency of the signal region}{cover rate of PSD-STK} for FCPs is approximately 86.0\%, as depicted in \added{panel} (d).}
\end{figure*}

The differences \added{in the charge distributions}  between MC protons and MC FCPs are depicted in Fig.~\ref{figs2}(a) and~\ref{figs2}(b). The integrals of MC FCPs are also drawn in the correspond\replaced{ing}{ence} panel to evaluate the selection of \added{the} signal region. \added{The signal region is defined as the area where the charge values of the PSD and STK are less than 0.84$e$ and 0.79$e$, respectively.}  The standard deviation $\sigma$ is obtained by dividing the full width at half maximum of the distributions by 2.35. The values corresponding to the signal region are obtained by adding $3\sigma$ to \added{the} peak value. The two-dimensional distributions of \added{the} PSD-STK charges of MC samples are depicted in Fig.~\ref{figs2}(c) and~\ref{figs2}(d) accompanied \replaced{by}{with} the signal region indicated by red lines. A combined integral efficiency of the signal region of up to 86\% is observed for MC FCPs, as shown in  Fig.~\ref{figs2}(d). The signal region is deemed to be \replaced{adequate for excluding}{suitable to exclude} the background\deleted{ radiation} \replaced{from}{of} proton MIP events, as depicted in Fig.~\ref{figs2}(c).

\begin{figure}[!htbp]
	\includegraphics[width=\linewidth]{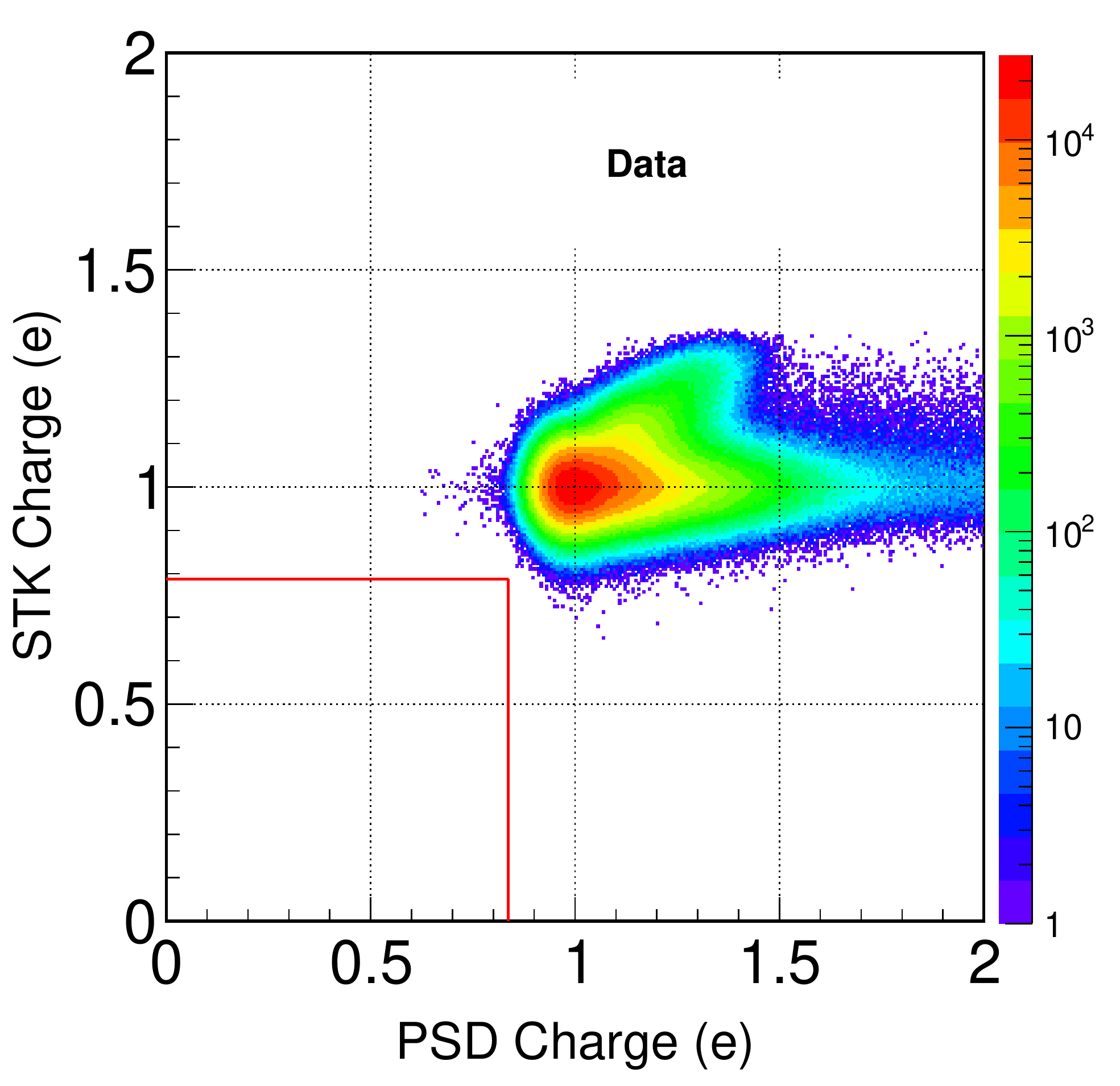}
	\caption{\label{fig2} The distribution of PSD-STK charge \replaced{for}{of} on-orbit data. The red lines indicate the signal region for FCP\added{s}. The signal region is \replaced{defined}{supposed} to cover candidate FCP event, while rejecting the proton background. No candidate event is observed to lie within the signal region. \added{The} portion above 1.1e of both PSD and STK charges corresponds to the events that inject from the bottom to the top of DAMPE. These events are low-energy secondary particles of \deleted{the} extensive air shower\added{s}.}
\end{figure}

Figure~\ref{fig2} shows the two-dimensional PSD-STK charge distribution of \added{the }on-orbit data, as well as the signal region \replaced{that}{which} is shown as the red lines. 

\section{\label{sec:level1}Results}
The flux of $\frac{2}{3}e$ FCPs is given by Eq.~\ref{eq1}
\begin{eqnarray}
	\Phi = \frac{N_{obs}}{T_{exp}\epsilon_{scale}\epsilon_{trig}A_{eff}\epsilon_{region}}
	\label{eq1}\replaced{,}{.}
\end{eqnarray}
where \deleted{the} $T_{exp}$ denotes the effective exposure time\added{ for this work}, $\epsilon_{scale}$ the pre-scale factor of MIP\replaced{T}{s}, $\epsilon_{trig}$ the efficiency of \added{the} \replaced{MIPT}{trigger}\added{ for FCPs}, $A_{eff}$ the effective acceptance\added{ for FCPs}, $\epsilon_{region}$ the efficiency of the signal region\added{ for FCPs}, and $N_{obs}$ the number of observed \added{FCPs }candidates. 
The results reported in this work are based on data recorded \replaced{from}{between} 01.01.2016 to 12.31.2020. The data are sampled from the latitude range, [-20$^{\circ}$, +20$^{\circ}$], where the $T_{exp}$ is equal to approximately  $2.34\times 10^7$ s, \replaced{followed by}{following} the deduction of the dead time of the detection system as well as the time \replaced{when}{during which} the satellite was in the South Atlantic Anomaly region. The \replaced{MIPT}{MIPs trigger} pattern\deleted{ of DAMPE  ~\cite{zhang2019design}} \added{is used} in this analysis, \replaced{which}{ 
Its trigger logic} relies on \replaced{simultaneous signals in}{ coincidence of signals corresponding to} the upper and lower layers of the BGO calorimete\added{r}. $\epsilon_{scale} = \frac{1}{4}$ is\added{ designed} for \replaced{MIPT}{MIPs trigger} and \deleted{the} $\epsilon_{trig} =  85.5\%$ is based on FCP simulation\added{s}.
 $A_{eff}$ for FCP is also estimated based on the MC sample as $A_{eff} = A_{geo} \times \epsilon_{sel}$, where \deleted{the} $A_{geo}$ denotes the geometr\replaced{ic}{y} acceptance\deleted{ of DAMPE}, \replaced{and}{which} is approximately equal to 3000 $\mathrm{cm^{2}sr}$, and \deleted{the} $\epsilon_{sel}$ denotes the total selection efficiency, which depends on the selection method. \deleted{The} $A_{eff} = 940~\mathrm{cm^{2}sr}$ is observed following the selection process. $\epsilon_{region}$ represents the efficiency of the signal region for FCPs and is evaluated to be 86\%. 
\replaced{Since}{While} no candidate event is observed\deleted{ to fall} within the signal region and the amount of background\deleted{ radiation} is negligible, for\deleted{ the determination of} the \deleted{flux }upper limit, $N_{obs}$ is taken to be 2.44 at \added{the} 90\% C. L. ~\cite{feldman1998unified}. 

We assume that the systematic uncertainties of FCPs are the same as those of singly charged MIP events. The combined systematic uncertainty $\delta$ includes the effects of trigger efficiency, track reconstruction, and the detection efficiency of \added{the} PSD and STK. 
\replaced{The systematic uncertainties are investigated in the selection procedure based on comparisons between the MC proton and on-orbit data.}{We investigated the systematic uncertainties of selection procedure based on comparisons between the MC proton and on-orbit data.} \replaced{A half}{Half} of the differences between the efficiencies corresponding to on-orbit data and MC protons are considered \replaced{as}{to be} systematic uncertainties\deleted{ of FCPs}. The total systematic uncertainty of the selections is given by
\begin{eqnarray}
	\delta = \sqrt{\delta_{trigger}^2 + \delta_{track}^2 + \delta_{charge}^2}
	\label{eq4}\added{,}
\end{eqnarray}
where $\delta_{trigger}$=1.1\%, $\delta_{track}$=2.9\%, and $\delta_{charge}$=0.5\% denote the corresponding systematic uncertainties of the trigger, track selection, and charge selection efficiencies, respectively. Systematic uncertainties corresponding to other very loose selections are negligible\added{, where the}\deleted{The} total uncertainty is 3.1\%. 

With systematic uncertaint\replaced{ies}{y is} considered, the flux upper limit of $\frac{2}{3}e$ FCP  is \replaced{found}{demonstrated} to be $\Phi<6.2 \times 10^{-10}~\mathrm{cm^{-2}sr^{-1} s^{-1}}$.

\begin{table*}
	\caption{\label{table1}The comparison between DAMPE and other similar types experiments.}
	\begin{ruledtabular}
		\begin{tabular}{lccl}
			Experiments &Geometr\replaced{ic}{y} acceptance($\mathrm{cm^2sr}$) &Exposure time\added{~}(s)&Upper limit\added{~}($\mathrm{cm^{-2}sr^{-1} s^{-1}}$)\\ \hline
			AMS-01 & 3000&$3.6\times 10^4$ &$3.0\times10^{-7}$ (95\%~C. L.) \\
			BESS & $1500$ & $3.2\times10^5$&$4.5\times10^{-7}$ (90\%~C. L.) \\
			DAMPE & $3000$ & $2.3\times10^7$&$6.2\times10^{-10}$ (90\%~C. L.) \\
		\end{tabular}
	\end{ruledtabular}
\end{table*}
\begin{figure}[!htb]
	\includegraphics[width=\linewidth]{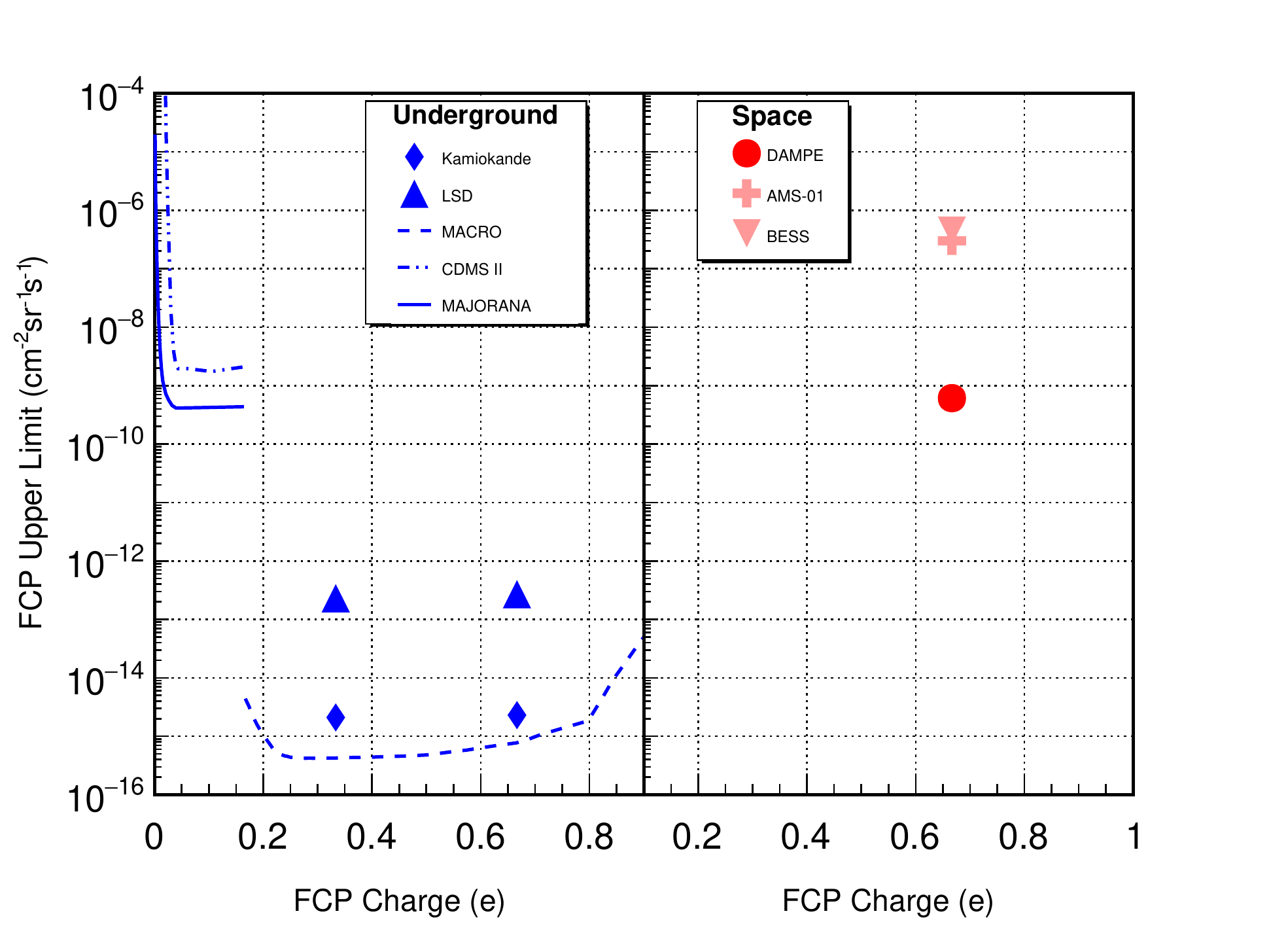}
	\caption{\label{fig3} FCP flux upper limit versus electric charge from different \replaced{cosmic ray experiments}{experiments of cosmic rays}. \added{The results of underground experiments which require the particles to have energy above $\sim$ 100s GeV are shown in the left panel. The results of space experiments  which detect the particles above a few GeV due to the limitation of geomagnetic cutoff are shown in the right panel.} \added{The }DAMPE \deleted{releases an }upper limit \deleted{at the 90\% C. L.} (red dot) \deleted{which }is lower than \replaced{those}{the results} from AMS-01 ~\cite{sbarra2003search} (light red \replaced{cross}{square}) and BESS ~\cite{fuke2008search} (\replaced{red}{green} inverted triangle). The results of the underground experiments such as LSD \cite{aglietta1994search} (\replaced{blue}{light green} triangle\added{s}), Kamiokande II \cite{mori1991search} (\replaced{blue}{light green} \replaced{full diamond}{star}), MACRO \cite{ambrosio2000search} (\replaced{blue}{light green} dashed line), CDMS II \cite{agnese2015first} (\replaced{blue}{light green} dotted line), and MAJORANA \cite{alvis2018first} (\replaced{blue}{light green} solid line) are shown also.}
\end{figure}

Table~\ref{table1} presents the complete results and some vital parameters of DAMPE, compared with other similar \deleted{type } experiments. Fig\replaced{ure}{.}~\ref{fig3}\deleted{3} shows \replaced{the upper limits from other}{some typical experiments for} FCP search\added{es}. Among underground \replaced{experiments}{apparatus}, MACRO yields the most sensitive upper limit. The CDMS II and MAJORANA \added{experiments} have high degrees of sensitivity to small charges because of the lower thresholds of the respective detection systems. Among space equipment, AMS-01 has \replaced{a large}{high} geometr\replaced{ic}{y} acceptance \cite{alcaraz1999search}, but \added{a} short exposure duration. BESS integrates \deleted{the }data gathered over four \deleted{times of }flight\added{s} to achieve a longer exposure time but its geometr\replaced{ic}{y} acceptance is limited. DAMPE has the longest and continuous exposure time as well as relatively large geometr\replaced{ic}{y} acceptance, and therefore it yields the most stringent FCP flux upper limit \replaced{for space experiments,}{in space,} \replaced{with an improvement of }{which is more stringent than the previous benchmark by} three orders of magnitude \added{over previous work}.

\section{\label{sec:level1}Summary}
\replaced{Based on on-orbit data obtained from DAMPE over a period of five years the results of the search for $\frac{2}{3}e$ FCPs in primary cosmic rays are as follows.}{The search for $\frac{2}{3}e$ FCPs in primary cosmic rays based on on-orbit data obtained from DAMPE over a period of five years is reported in this study.} \replaced{No FCP signals are observed and a}{FCP MIP events are searched, but no candidate is observed. The} flux upper limit \replaced{of}{is determined to be} $\Phi<6.2 \times 10^{-10}~\mathrm{cm^{-2}sr^{-1} s^{-1}}$ \added{is established} at \added{the} 90\% C. L.
\replaced{A precise measurement of the flux or a conservative flux upper limit}{Precise and prudent measurement of the flux or the flux upper limit} is essential to \replaced{construct and constrain the model}{detect the existence} of FCPs. \deleted{Extensive experimentation is required to enhance the comprehension of the modelling and production of FCPs. }Most of the previously performed underground experiments assumed that FCPs would exhibit long penetration paths, which, in turn, requires them to have energy exceeding a few hundred GeV. \replaced{Given the effective energy threshold arising from the geometric cutoff}{Considering energy threshold of the geomagnetic cutoff}, experiments in space can be used to detect FCPs with energy as low as a few GeV. \deleted{As an apparatus on orbit, }DAMPE \deleted{can }serve\added{s} as a novel observation platform and enable\added{s} \added{a} long-term, continuous search for relatively low-energy FCPs in primary cosmic rays. In the future, with the accumulation of more on-orbit data, DAMPE is expected to \replaced{perform even}{enable} more sensitive FCP searches.

\begin{acknowledgments}
	
	We acknowledge Prof. Xiaowei Tang and Prof. Zhengguo Zhao for their suggestions and encouragement for this work. \added{We also thank Mr. Wilson J. Huang for proofreading the manuscript.}
	The DAMPE mission was funded by the strategic priority science and technology projects in space science of Chinese Academy of Sciences. 
	In China the data analysis is supported by the National Key Research and Development Program of China (No. 2016YFA0400200), the National Natural Science Foundation of China (No. 11673021, No. U1738205, No. U1738208, No. U1738139, No. U1738135, No. 11705197, No. 11851302), 
	the strategic priority science and technology projects of Chinese Academy of Sciences (No. XDA15051100), the Youth Innovation Promotion Association CAS (Grant No. 2021450), the Outstanding Youth Science Foundation of NSFC (No. 12022503). 
	In Europe the activities and data analysis are supported by the Swiss National Science Foundation (SNSF), Switzerland, the National Institute for Nuclear Physics (INFN), Italy, and the European Research Council (ERC) under the European Union’s Horizon 2020 research and innovation programme (No. 851103).
\end{acknowledgments}

\nocite{*}

\bibliography{apssamp}

\end{document}